\documentclass{article} 
\usepackage{main,times}

\usepackage{amsmath,amsfonts,bm}

\def\eqref#1{equation~\ref{#1}}

\def\1{\bm{1}}

\DeclareMathAlphabet{\mathsfit}{\encodingdefault}{\sfdefault}{m}{sl}
\SetMathAlphabet{\mathsfit}{bold}{\encodingdefault}{\sfdefault}{bx}{n}

\usepackage{hyperref}
\usepackage{url}
\usepackage[none]{hyphenat}

\usepackage{tabularx,booktabs,makecell,array}
\newcolumntype{Y}{>{\raggedright\arraybackslash}X}

\usepackage{graphicx}   
\usepackage{subcaption} 

\usepackage[margin=1in]{geometry}
\usepackage[T1]{fontenc}
\usepackage[utf8]{inputenc}
\usepackage{hyperref}
\usepackage{tabularx}
\usepackage{booktabs}
\usepackage{longtable}
\usepackage{array}
\usepackage{enumitem}
\usepackage{caption}
\usepackage{verbatim}
\usepackage{amsmath}
\usepackage{amssymb}

\newcolumntype{L}{>{\raggedright\arraybackslash}X}

\usepackage{fvextra} 
\RecustomVerbatimEnvironment{verbatim}{Verbatim}{
  breaklines,          
  breakanywhere,       
  breaksymbolleft={},  
  breakindent=0pt      
}

\usepackage{array}
\usepackage{longtable}
\usepackage{tabularx}
\usepackage{booktabs}
\usepackage{verbatim}
\usepackage{geometry}

\usepackage{lscape}
\usepackage{graphicx}

\usepackage{fvextra}

\title{\sloppy\nohyphens{SWE-Bench++: A Framework for the Scalable Generation of Software Engineering Benchmarks from Open-Source Repositories}}

\author{
  \parbox{\linewidth}{\centering
    \vspace{2em}
    Lilin Wang, Lucas Ramalho, Alan Celestino, Phuc Anthony Pham, Yu Liu, \\
    \textbf{Umang Kumar Sinha, Andres Portillo, Onassis Osunwa, Gabriel Maduekwe} \\[0.5em] 
    \textbf{Research \& Development, Turing} \\[0.3em] 
    \texttt{}
  }
}

\usepackage[most]{tcolorbox}
\usepackage{listings}
\usepackage{xcolor}

\lstdefinelanguage{json}{
    basicstyle=\ttfamily\small,
    columns=fullflexible,
    showstringspaces=false,
    commentstyle=\color{gray},
    keywordstyle=\color{blue},
    stringstyle=\color{teal},
    breaklines=true,
}

\lstdefinelanguage{logoutput}{
    basicstyle=\ttfamily\small\color{black!80},
    backgroundcolor=\color{gray!10}, 
    frame=none,
    breaklines=true,
    columns=fullflexible
}

\newtcolorbox{promptbox}[2][]{
    enhanced,
    breakable,
    title={#2},
    colback=gray!5, 
    colframe=black!75, 
    coltitle=white,
    fonttitle=\bfseries,
    attach boxed title to top left={xshift=5mm, yshift*=-3mm},
    boxed title style={colback=black!75},
    boxrule=0.5mm,
    top=5mm, 
    #1
}

\lstdefinestyle{jsonWithComments}{
    language=json,
    morecomment=[l]{//},     
    commentstyle=\color{gray}\ttfamily,
    frame=none,
    backgroundcolor=\color{white}
}
\begin{document}

\maketitle

\vspace{-36pt}
\definecolor{LightBlue}{rgb}{0.35,0.55,1} 
\begin{center}
\small\textbf{Project Page: }\href{https://research.turing.com/swebench}{\textbf{\textcolor{LightBlue}{https://research.turing.com/swebench}}}
\end{center}

\begin{abstract}
Benchmarks like SWE-bench have standardized the evaluation of Large Language Models (LLMs) on repository-level software engineering tasks. However, these efforts remain limited by manual curation, static datasets, and a focus on Python-based bug fixes. We introduce SWE-Bench++, an automated framework that generates repository-level coding tasks from open-source GitHub projects. Unlike synthetic approaches, our pipeline harvests live pull requests to cover both bug fixes and feature requests across 11 languages from open-source GitHub repositories. SWE-Bench++ turns GitHub pull requests (PRs) into reproducible, execution-based tasks via four stages: programmatic sourcing, environment synthesis, test oracle extraction, and quality assurance. A final hint-guided trajectory synthesis step converts instances that strong models fail to solve into training trajectories. Our initial benchmark consists of 11,133 instances from 3,971 repositories across 11 languages. On a subset of 1,782 instances of this benchmark, today's strongest models perform as follows: \texttt{claude-sonnet-4.5} achieves 36.20\% pass@10, \texttt{gpt-5-2025-08-07} 34.57\%, \texttt{gemini/gemini-2.5-pro} 24.92\%, and \texttt{gpt-4o} 16.89\%. We further demonstrate the utility of our dataset by showing that fine-tuning on SWE-Bench++ instances yields measurable improvements on the SWE-bench Multilingual benchmark. SWE-Bench++ provides a scalable, multilingual benchmark for evaluating and improving repository-level code generation.

\end{abstract}

\vspace{-10pt} 
\begin{figure}[h!]
  \centering
  \includegraphics[width=\textwidth]{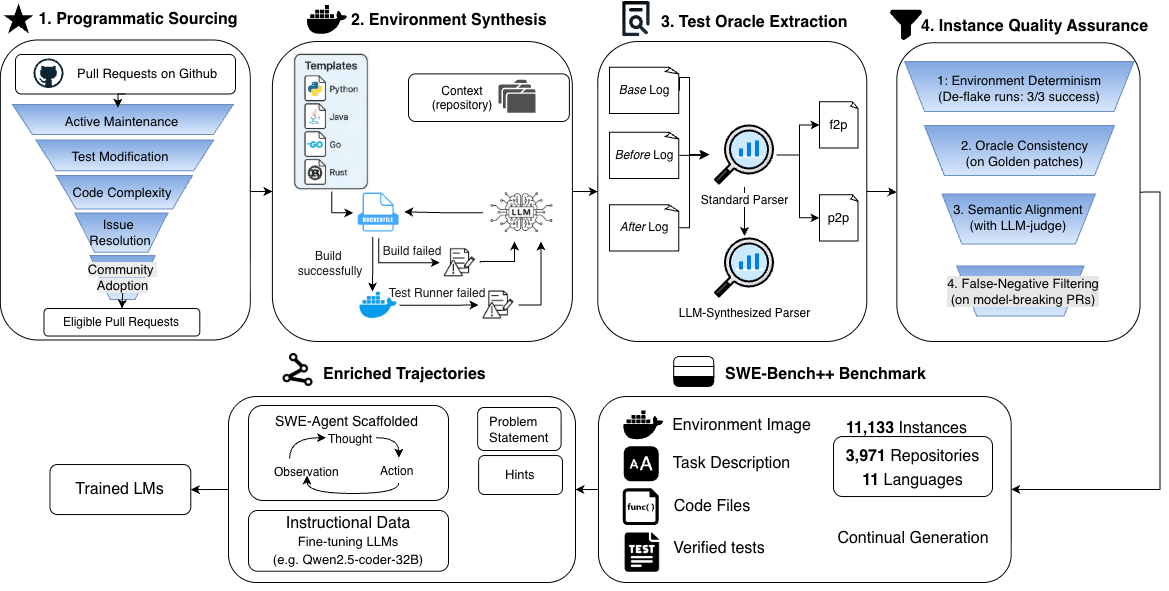}
  \caption{The SWE-Bench++ Framework. Unlike static benchmarks, our pipeline uses constrained neural synthesis to generate reproducible Docker environments and adaptive log parsers across 11 languages. A three‑state, state‑differential oracle automatically classifies tasks as bug fixes or feature requests, producing verified benchmark instances and hint‑guided training trajectories at scale.}
  \label{fig:framework}
\end{figure}
\vspace{-10pt} 

\section{Introduction}

The evaluation of Large Language Models (LLMs) coding agents has shifted from isolated function synthesis (e.g., HumanEval \citep{chen2021codexhumaneval}) to repository-level software engineering. SWE-bench introduced a repository-level benchmark based on real GitHub issues, enabling more realistic evaluation of LLM coding agents \citep{jimenez2024swebench}. However, it relies on manual curation and covers only 12 Python repositories. This scale is too small to capture the structural and linguistic diversity of open-source projects.

Recent efforts to improve scalability have followed two paths. First, benchmarks such as Multi-SWE-bench \citep{zan2025multiswebench} and SWE-bench Multilingual \citep{yang2025swesmithscalingdatasoftware} extend the evaluation framework to languages like Java and Rust, but rely on manual curation and cover only a few dozen repositories. Second, Python-centric automation such as SWEE-bench \citep{vergopoulos2025-automated-generation-repository-level-coding-tasks-icml} scales to hundreds of repositories, but remains restricted to Python and suffers from two technical limitations. First, its two-state test oracle (Before-patch → After-patch) is not designed to extract feature requests that introduce new APIs or functionality, as these cause the Before state to fail to build due to missing symbols—cases that existing pipelines must filter out as errors. This methodological constraint restricts automated benchmarks to scenarios where tests can be executed in both states, severely limiting the coverage of feature requests. Second, it relies on static regular expressions for log parsing, preventing it from scaling to the "long tail" of repositories with heterogeneous test runners and non-standard outputs.

Furthermore, recent works like \textbf{SWE-Smith} \citep{yang2025swesmithscalingdatasoftware} and \textbf{SWE-Flow} \citep{zhang2025sweflow} have attempted to scale data generation via synthetic means, such as synthesizing tasks from Test-Driven Development (TDD) patterns. While valuable for training, these synthetic approaches are less well suited for evaluating models on “in-the-wild” distributions. They lack the noisy, complex, and historical nature of human-written code. Additionally, the static nature of all aforementioned benchmarks introduces a critical \textbf{data contamination risk}: most instances were created before the training cutoff of modern models, rendering them prone to memorization.

To bridge these gaps, we present \textbf{SWE-Bench++}, an automated multilingual framework that generates software engineering benchmarks from GitHub pull requests. Unlike previous approaches, our methodology provides a systematic pipeline that transforms raw GitHub repositories into executable evaluation environments without human intervention.

To address these challenges, we introduce three mechanisms:

\begin{enumerate}
  \item \textbf{Constrained Environment \& Oracle Synthesis:} Existing frameworks often rely on unstructured command extraction for environments and static regexes for logs, which can be brittle at scale. We introduce synthesis engines for both infrastructure and verification. For environments, we use template-guided synthesis to populate security-hardened Dockerfiles (e.g., enforcing multi-stage builds). This approach combines LLM reasoning with static templates, achieving approximately 137\% higher yield on Python repositories than a \emph{SetUpAgent} baseline when both are run on the same pool of 2,377 valid pull requests. For verification, we use adaptive parser synthesis to generate custom Python parsers for heterogeneous logs, when deterministic parsers fail. This constrained neural synthesis approach standardizes environment and oracle construction across 11 languages and 3,971 repositories.

  \item \textbf{State-Differential Task Classification:} Current benchmarks struggle to distinguish between bug fixes and feature requests, often discarding instances where the pre-fix codebase fails to build. We implement a state-differential oracle that compares three repository states: \emph{Base}, \emph{Before} (test patch applied), and \emph{After} (full PR applied). We treat specific build failures in the \emph{Before} state not as errors, but as semantic signals for Feature Requests (where tests rely on yet-to-be-implemented code). This allows us to verify both regression fixes and new feature implementations.

  \item \textbf{Hint-Guided Trajectory Synthesis:} Standard training data generation (e.g., SWE-Gym) relies on passive filtering of easy tasks that agents can already solve. We introduce an active Hint Injection Algorithm that converts model-breaking instances (where SOTA models fail) in SWE-Bench++ environments into executable training trajectories. By injecting function signatures and dependency graphs as hints, we scaffold the agent to solve previously impossible tasks. Fine-tuning on just 145 of these trajectories improves cross-lingual performance from 1.6\% to 3.6\% on SWE-bench Multilingual. 

\end{enumerate}

\noindent\textbf{Contributions:} Our work makes the following contributions:
\begin{itemize}
  \item \textbf{Large-scale benchmark:} We construct 11,133 repository-level instances from 3,971 repositories, covering diverse build systems and coding patterns.
  \item \textbf{Automated multilingual environments:} Our pipeline automatically synthesizes Docker environments and log parsers across 11 languages.
  \item \textbf{Broader task coverage:} Our state-differential oracle identifies both bug fixes and feature requests, increasing feature-like coverage compared to prior benchmarks (e.g., 9\% in SWE-bench). 
  \item \textbf{Contamination‑Aware Evaluation}: SWE‑Bench++ is constructed from dated GitHub pull requests and can be filtered by PR creation date, enabling temporally separated evaluation sets that reduce data‑contamination risk for future models.
\end{itemize}

Table 1 compares SWE-Bench++ with prior benchmarks and frameworks.

\begin{table}[t!] 
    \centering
    \scriptsize 
    \setlength{\tabcolsep}{1.5pt} 
    \renewcommand{\arraystretch}{1.1} 
    \caption{Comparison of software engineering benchmarks and frameworks.}
    \label{tab:benchmark_compare}
    
    \begin{tabularx}{\textwidth}{
        >{\raggedright\arraybackslash\hsize=0.7\hsize}X  
        >{\raggedright\arraybackslash\hsize=0.96\hsize}X 
        >{\raggedright\arraybackslash\hsize=0.96\hsize}X 
        >{\raggedright\arraybackslash\hsize=0.96\hsize}X 
        >{\raggedright\arraybackslash\hsize=0.96\hsize}X 
        >{\raggedright\arraybackslash\hsize=0.96\hsize}X 
        >{\raggedright\arraybackslash\hsize=0.96\hsize}X 
        >{\raggedright\arraybackslash\bfseries\hsize=1.54\hsize}X 
    }
        \toprule
        \textbf{Feature} & 
        \textbf{SWE-bench / Multi-SWE} & 
        \textbf{SWEE-bench (SetUpAgent)} & 
        \textbf{SWE-Smith} & 
        \textbf{SWE-Flow} & 
        \textbf{SWE-Fixer} & 
        \textbf{SWE-Gym} & 
        \textbf{SWE-Bench++ (Ours)} \\
        \midrule
        
        \textbf{Function} & 
        Benchmark Dataset & 
        Benchmark Generator & 
        Syn. Data Gen. & 
        Syn. Data Gen. & 
        Solver Tool & 
        RL Interface & 
        Live Benchmark Generator \\
        \midrule
        
        \textbf{Generation} & 
        Manual Curation & 
        Automated & 
        Synthetic & 
        Synthetic & 
        Static Scrape (PRs only) & 
        N/A & 
        Automated \\
        \midrule
        
        \textbf{Gen. Scope} & 
        N/A & 
        Container & 
        Bug-fix pairs & 
        Fix-test pairs & 
        N/A & 
        N/A & 
        Container, Log Parser, Trajectory \\
        \midrule
        
        \textbf{Env Strategy} & 
        Pre-build Images & 
        Extract cmds & 
        Pre-build image & 
        Pre-verified images & 
        N/A & 
        Pre-build images & 
        Auto-Synthesized \\
        \midrule
        
        \textbf{Scale} & 
        12 / 42 & 
        514 & 
        128 & 
        74 & 
        $\sim$110,000 & 
        358 & 
        3,971 \\
        \midrule
        
        \textbf{Languages} & 
        Python / 9 & 
        Python Only & 
        Python Only & 
        Python Only & 
        Python Centric & 
        N/A & 
        11 (Automated) \\
        \midrule
        
        \textbf{Task Scope} & 
        Bug Fixes & 
        Bug Fixes & 
        Bug Fixes Only & 
        TDD Incr. Dev. & 
        Simple Bugs & 
        Bug Fixes & 
        Bugs \& Feature Requests \\
        \midrule
        
        \textbf{Log Parsing} & 
        Static Regex & 
        Static Regex & 
        Static Regex & 
        Static Regex & 
        N/A & 
        N/A & 
        Syn. Adaptive Parsers \\
        \midrule
        
        \textbf{Distribution} & 
        Organic & 
        Organic & 
        Synthetic & 
        Synthetic & 
        Organic & 
        Organic & 
        Organic \\
        \midrule
        
        \textbf{Freshness} & 
        Static & 
        Static & 
        N/A & 
        N/A & 
        Static & 
        Static & 
        Continuous (Living) \\
        \bottomrule
    \end{tabularx}%
\end{table}

\section{Related Work}

We review related work primarily through the lens of the structural limitations our benchmark aims to address. While \textbf{SWE-bench} and its manually curated variant \textbf{SWE-bench Verified} \citep{chowdhury2024swebenchverified} established the gold standard for evaluating LLMs on repository-level tasks, their static and manually intensive nature creates fundamental barriers to scaling software engineering evaluation.

\paragraph{Scalability}
First, synthetic generation approaches like \textbf{SWE-Smith} and \textbf{SWE-Flow} utilize LLMs to \emph{synthesize} training signals---either by injecting bugs into existing codebases or by inferring incremental steps from Test-Driven Development (TDD) patterns. While valuable for training efficiency, these synthetic tasks often lack the noise, ambiguity, and ``in-the-wild'' distribution of real human-written code. Second, static data scaling efforts like \textbf{SWE-Fixer}~\citep{xie2025swefixer} aggregate massive datasets by scraping GitHub history. However, it operates as a retrieval-based pipeline without execution environments, prioritizing raw volume over execution-based verification. Third, compute scaling frameworks like \textbf{SWE-Gym}~\citep{pan2025swegym} transform existing benchmarks into reinforcement learning (RL) environments to generate millions of agent trajectories. While this scales experience, it remains bound to the limited problem set of the original manually curated datasets. Finally, attempts to automate organic task collection, such as \textbf{SWEE-bench} \citep{vergopoulos2025-automated-generation-repository-level-coding-tasks-icml}, utilize agents (e.g., \emph{SetUpAgent}) to scaffold environments but remain restricted to Python and lack support for feature requests.

\paragraph{Data Contamination and ``Live'' Evaluation}
Static benchmarks are highly vulnerable to data contamination, as instances created before a model's knowledge cutoff are frequently memorized during pre-training. The community has responded with ``live'' benchmarks such as \textbf{SWE-bench-Live} and \textbf{LiveCodeBench}, which continuously harvest new problems \citep{zhang2025swebenchgoeslive, jain2025-livecodebench-holistic-contamination-free-large-livecodebench}. Similarly, \textbf{SWE-bench Pro} attempts to mitigate contamination by incorporating private, commercial repositories \citep{deng2025swebenchpro}. However, these solutions often rely on specific language ecosystems or lack the fully automated, multi-stage verification pipeline required to scale beyond hundreds of tasks to thousands.

\paragraph{The Weak Test Oracle Problem}
Standard evaluation protocols assume that passing a developer-written test suite equates to a correct fix. However, this ``test oracle'' is often unreliable. Empirical studies on \textbf{SWE-bench} have revealed that a significant percentage of plausible patches---those that pass the provided tests---are semantically incorrect or diverge from the ground truth. This oracle limitation can lead to overestimation of model capabilities, highlighting the need for more rigorous, state-based verification methods.

\paragraph{Environment Reproducibility Challenges}
Accurately reconstructing historical development environments is a primary scalability bottleneck. Complex dependency trees and version mismatches often lead to ``environment rot,'' where test failures result from configuration errors rather than faulty model patches. While agentic systems like \textbf{SetUpAgent} \citep{vergopoulos2025-automated-generation-repository-level-coding-tasks-icml} automate some aspects of environment creation through command extraction from documentation, they often lack the reliability of deterministic, template-guided scaffolding, restricting their success rate across polyglot repositories.

\paragraph{Solution Leakage and Ambiguity}
Finally, the quality of problem statements in existing benchmarks varies significantly. A recent analysis by \textbf{SWE-bench+} ~\citep{aleithan2024swebenchplus} revealed that \textbf{32.67\%} of successful agent resolutions involved ``solution leakage,'' where the correct code or a direct pointer was explicitly present in the issue description or comments. This allows agents to ``solve'' tasks via information retrieval rather than reasoning, further skewing leaderboard rankings.

\section{Methodology: The SWE-Bench++ Framework}

SWE-Bench++ is a four-stage pipeline (Figure 1) that converts GitHub pull requests into executable software engineering tasks.

\subsection{Stage 1: Programmatic Sourcing}
The pipeline begins with a broad search to identify candidate tasks that represent realistic software maintenance and evolution. We employ scalable filters to process the GitHub firehose, selecting repositories and pull requests (PRs) that meet the following criteria: (a) active maintenance histories with recent commit activity; (b) evidence of community adoption (e.g., $>100$ stars) and a recognizable testing framework; (c) substantial complexity, defined by codebases exceeding 10k lines of code; (d) merged PRs that explicitly resolve an issue (ensuring a link between a natural language problem description and a coding solution); and (e) PRs that include edits or additions to test files. This coarse filtering casts a wide net, identifying millions of potential candidates for the compute-intensive stages that follow.

\subsection{Stage 2: Environment Synthesis}
Once a high-quality PR is identified, we create a reproducible execution environment that mirrors the repository at the time the issue was present. We treat environment generation as a constrained synthesis problem. Rather than relying on unstructured generation, our system employs a hybrid architecture: parameterized Dockerfile templates enforce structural validity and security standards (e.g., multi-stage builds), while an LLM infers missing dynamic dependencies and versions (e.g., package versions, entry points) that cannot be parsed statically. 

\subsubsection{Template-Based Scaffolding}
To mitigate the security risks and logical errors inherent in generating Dockerfiles from scratch, our system utilizes a library of vetted, language-specific templates (e.g., Python, Java, Go, Rust). These templates enforce best practices, such as multi-stage builds, minimal base images, and non-root user execution, providing a secure structural skeleton. The templates contain semantic placeholders for dynamic content, including language versions, dependency installation commands, and entry points. The agent's objective is to populate these placeholders, thereby combining the robustness of a human-engineered structure with the flexibility of an LLM.

\subsubsection{Iterative refinement with build-and-test feedback}
The instantiation process is an iterative loop that uses Docker builds and test runs as validation signals. The process begins with Repository Analysis and Plan Generation, where the synthesis engine performs deep structural analysis rather than heuristic approaches that simply scrape documentation files (e.g., \texttt{README}). The LLM is granted controlled tool access via a Model Context Protocol (MCP) server that exposes repository-level operations (\texttt{clone}, \texttt{list}, \texttt{read}). Through these deterministic calls, the synthesis engine materializes the target PR at the correct commit, traverses the project tree, and inspects build scripts and manifests verbatim. This evidence is incorporated into a structured JSON plan. This tool-augmented introspection enables precise identification of required base images, package managers, and entry points, reducing hallucinations compared to static text extraction.

Following this analysis, the system enters a Build-Feedback Loop. It injects the JSON values into the template and attempts a \texttt{Docker build}. If the build fails, the synthesis engine captures the standard error (\texttt{stderr}) output. This error trace serves as feedback for the LLM, which generates a corrected JSON plan. This self-correction loop continues until success or a retry limit (set to 5) is reached. Crucially, a successful build guarantees only syntactic validity, so the final phase is
the TestRun-Feedback Loop. Here, the system spins up the container to verify that the test runner executes correctly, ensuring that no tests fail due to environment misconfiguration before advancing the instance to the next stage.

We quantify the impact of this template‑guided design in Section 4.1, where we show that substituting a SetUpAgent‑style baseline reduces yield on Python repositories to $\approx 40\%$ of our full system.

\subsection{Stage 3: Automated State-Differential Test Oracle Extraction}
Once the environment is ready, this stage executes the code across multiple repository states and extracts the \emph{test oracle} through automated log parsing.

\subsubsection{Three Repository States}
We consider three snapshots of the repository relative to the PR: \textbf{base} (parent commit of the PR), \textbf{before} (base plus test-file changes only), and \textbf{after} (full PR, including implementation changes). The pipeline runs all tests under these three states---Base, Before, and After---producing three distinct execution logs.

\subsubsection{State-Differential Classification Logic}
Existing benchmarks primarily support \textbf{Scenario A: Regression / Bug Fix}---they assume that the \texttt{Before} state is buildable and focus solely on ``Fail-to-Pass'' (F2P) tests. However, based on observations from large-scale real-world datasets, we find that this definition fails to capture \textbf{Scenario B: Feature Request}.

To address this limitation, we broaden the definition of F2P to encompass both scenarios, allowing the framework to handle regression fixes and feature additions in a unified manner.
\begin{itemize}
  \item \textbf{Scenario A: Regression / Bug Fix.} If the \texttt{Before} state builds successfully, we execute the modified tests.
  \begin{itemize}
    \item \emph{F2P:} Tests that fail in \texttt{Before} and pass in \texttt{After}. These represent the regression.
    \item \emph{P2P:} Tests that pass in both. These represent the constraint to not break existing functionality.
  \end{itemize}
  \item \textbf{Scenario B: Feature Request.} If the \texttt{Before} state fails to build (due to missing symbols/dependencies introduced in the PR):
  \begin{itemize}
    \item \emph{F2P:} We identify newly added tests in the PR that pass in the \texttt{After} state. The build failure in \texttt{Before} serves as the confirmation that the feature was absent.
  \end{itemize}
\end{itemize}

Section 4.1 reports an ablation showing that disabling adaptive parser synthesis reduces the final dataset by about 16\% (to $\approx 84\%$ of its original size).

\subsubsection{Synthesized Adaptive Log Parsing}
To automatically derive the test oracle---structured evidence of which tests pass or fail---the system employs a hybrid architecture designed to handle diverse test frameworks and noisy outputs.

We implement a hierarchical parsing strategy. The system first attempts deterministic symbolic parsing (using high-precision regex) for standard frameworks (full list in Appendix \ref{sec:log-parsers}) to ensure zero-cost accuracy. When symbolic parsing fails (e.g., unrecognized formats), the system falls back to neural synthesis, where an LLM generates a custom Python parser. To verify the generated parser, we use synthetic failure injection. An LLM modifies assertions in the code to force a failure; if the parser correctly identifies this failure, it is considered valid.

\begin{enumerate}
  \item \textbf{Self‑correcting synthesis loop:} The system executes the synthesized parser on sample logs and feeds any crashes or implausible test counts back to the LLM, iteratively refining the parser. This ensures resilience against noisy or non-standard output formats.

  \item \textbf{Synthetic failure injection (Balanced Evaluation):} We inject artificial failing assertions and check that the parser flips the corresponding test outcomes, providing an empirical check that it discriminates \texttt{PASS} vs \texttt{FAIL} correctly.

\end{enumerate}

\subsection{Stage 4: Automated Instance Quality Assurance \& Verification}
To ensure the reliability of the generated instances, we deploy a four-layer \textbf{Automated Quality Assurance (AutoQA)} pipeline designed to identify and reject unstable environments, flaky tests, and ambiguous problem statements.

\paragraph{Layer 1: Environment Determinism (Build Stability)}
Each dockerized environment is built and instantiated three times, and we retain only instances where the testbed initializes successfully in all three runs, filtering out flaky build dependencies.

\paragraph{Layer 2: Oracle Consistency (Test Determinism)}
To confirm that the test suite yields identical outcomes across runs, we validate Test Determinism. We execute the ``golden solution'' (the ground-truth patch) against the verify/regression tests three times in independent containers. We retain only those instances where the test results (Pass/Fail) are identical across all three runs, eliminating ``flaky'' tests that pass or fail based on timing conditions.

\paragraph{Layer 3: Semantic Alignment \& Automated Curation} 
We employ a rubric-based LLM-Judge \ref{app:llm-judge} to evaluate the alignment between the problem statement and the test oracle. Empirically, these LLM-based reviewers achieve precision close to senior human annotators; see Appendix C.3 (Table~\ref{tab:auto-reviewers}) for a detailed comparison of automated reviewers and human raters. While instances with fundamental ambiguity and misalignment are rejected ("Low Quality"), we identify a recoverable class of "Medium Quality" instances where tests rely on implementation details not explicitly requested in the issue (e.g., new accessor methods). For these, we trigger an Automated Curation module: the system analyzes the code patch to extract the signatures of implicit dependencies and appends them to the problem statement as "Hints". This systematically repairs underspecified tasks, transforming them into high-fidelity, verifiable instances without manual per‑instance curation. Details can be found in Appendix \ref{app:hints}.

\paragraph{Layer 4: False-Negative Filtering (model-breaking verification)}
To ensure that our benchmark accurately measures the upper bounds of model capability, we perform deep inspection on instances where state-of-the-art (SOTA) models fail to generate a solution. The goal is to distinguish between True Negatives (model limitation) and False Negatives (dataset defects). We utilize an automated \textbf{trajectory \& log inspection} module that parses the execution trace of failed SOTA attempts. Instances where the model failure stems from infrastructure artifacts (e.g., unsupported tool, tool crashes), unsupported dependencies, or underspecified problem statements are flagged and removed. This ensures that the remaining ``hard'' instances represent genuine reasoning challenges.

\paragraph{Human Verification (Verified Subset)}
While the core pipeline is fully automated, we established a manually verified subset for high-precision evaluation. We recruited 82 pre-screened annotators to conduct comprehensive manual verification on the model-breaking instances retained from Layer~4, following the guidelines of SWE-bench Verified.

\subsection{Application: Hint‑Guided Trajectory Synthesis for Training}

Stages 1–4 fully specify the SWE‑Bench++ benchmark and its evaluation pipeline. In this section, we illustrate an application of these environments: using a hint‑guided curation algorithm to convert model‑breaking instances into high‑fidelity training trajectories. 

Unlike frameworks such as SWE-Gym \citep{pan2025swegym}, which generate data by passively filtering for trajectories where an agent naturally succeeds, we target model-breaking instances that SOTA models fail to solve. To convert these failures into valuable training signals, we introduce a \textit{Hint-Guided Curation} algorithm:

\begin{enumerate}
    \item \textbf{Failure Identification}: The system identifies instances where SOTA baselines consistently fail ($0\%$ pass rate)
    
    \item \textbf{Contextual Scaffolding}: The pipeline analyzes the ground-truth patch to extract critical function signatures and dependency graphs. These are injected as ``Hints'' into the agent's system prompt to bound search space.
    
    \item \textbf{Guided Resolution}: The agent retries the task with this scaffold. This mechanism raises the pass rate on these ``hard'' tasks from $0\%$ to $\sim70\%$, harvesting successful reasoning traces on problems that were previously unsolvable.
    
    \item \textbf{Contamination Control \& Outcome}: To prevent the model from learning to rely on hints, we apply a \textit{Thought Regeneration} pass: an LLM rewrites the agent's reasoning trace to exclude hint-related keywords while preserving the logical solution path. This produces a dataset of ``frontier'' trajectories---instances at the exact boundary of capability---providing significantly higher information gain for fine-tuning than the ``easy'' instances captured by passive filtering.
\end{enumerate}

\section{Empirical Validation}
We evaluate SWE-Bench++ along four axes: (i) pipeline yield and dataset properties (Section 4.1), (ii) agent performance baselines (Section 4.2), (iii) fine-tuning experiments (Section 4.3), and (iv) qualitative failure analysis (Appendix H).

\subsection{EVALUATING THE SWE-Bench++ PIPELINE AND DATASET}
\label{sec:pipeline-analysis}

\paragraph{Yield \& Throughput Analysis}
We initialized the sourcing module with \textbf{137,048} candidate Pull Requests (PRs) meeting our activity and complexity criteria. Of these, \textbf{28,513 (20.8\%)} successfully passed the Environment Synthesis stage and State-Differential Test Oracle Extraction (Stage 2 and Stage 3), yielding a reproducible Docker container with successfully parsed logs. This success rate varies by language ecosystem, with \textbf{Python (41\%)} and \textbf{Java (27\%)} showing the higher resilience, while compiled languages like \textbf{C++ (9.5\%)} present greater environmental challenges due to complex toolchain dependencies. Following the Automated Quality Assurance pipeline (Stage 4), \textbf{39\%} of the dockerized instances were verified as deterministic and aligned, resulting in a final dataset of \textbf{11,133 instances}. The end-to-end processing time averages around 67 minutes per instance, dominated by the compilation and testing latency of the Docker build process.

Controlled comparison with SetUpAgent: To isolate the effect of template‑guided environment synthesis, we re‑ran a SetUpAgent pipeline and our method on the same pool of 2,377 valid Python pull requests originally selected by SWEE‑bench. Our approach successfully recovers deterministic, dockerized environments for \textasciitilde2.37× as many PRs (a \textasciitilde137\% relative increase in yield) as the SetUpAgent baseline, under identical success criteria.

\paragraph{Ablation: impact of template‑guided environment synthesis}
On Python repositories, replacing our template‑guided Dockerfile synthesis with a SetUpAgent‑style command‑extraction approach reduces the number of successfully dockerized PRs to $\approx 40\%$ of our full system’s yield in Python repositories. Combined with the controlled comparison above, this suggests that the majority of our yield gains over SetUpAgent come from constraining generation within vetted templates.

\paragraph{Ablation: impact of adaptive parser synthesis} 
If we disable the adaptive log‑parser synthesis stage and rely exclusively on deterministic regex parsers, the number of instances that survive the full pipeline drops to $\approx 84\%$ of the final 11,133. The missing 16\% primarily correspond to repositories with non‑standard or noisy test outputs that our fixed parsers cannot reliably interpret, highlighting the importance of the neural parser synthesis stage.

\begin{table}[ht]
    \centering
    \caption{Dockerization success rates (Yield) by language}
    \label{tab:language_yield}
    
    \resizebox{\textwidth}{!}{%
        \begin{tabular}{lccccccccccc}
            \toprule
            \textbf{Language} & Python & Go & TS & JS & Ruby & PHP & Java & Rust & C++ & C\# & C \\
            \midrule
            \textbf{Yield} & 41.0\% & 41.0\% & 40.0\% & 39.0\% & 38.0\% & 38.0\% & 27.0\% & 19.0\% & 11.0\% & 10.0\% & 9.5\% \\
            \bottomrule
        \end{tabular}%
    }
\end{table}

\begin{table}[t]
  \caption{Yield analysis across processing stages}
  \label{tab:yield_analysis}
  \centering
  \begin{small}
  \begin{tabular}{lrrp{0.38\linewidth}}
    \toprule
    \textbf{Stage} & \textbf{Count} & \textbf{Yield} & \textbf{Notes} \\
    \midrule
    Stage 1 (Sourcing) & 137{,}048 & 100\% & Filtered by activity \& complexity, etc. \\
    Stage 2 \& 3 (Environment Synthesis \& Log Parsing) & 28{,}513 & 20.8\% & High variance: Python (41\%) vs.\ C++ (9.5\%). \\
    Stage 4 (Quality Assurance) & 11{,}133 & 8.1\% & High-fidelity deterministic instances. \\
    \bottomrule
  \end{tabular}
  \end{small}
\end{table}

\begin{figure}[htbp]
  \centering
  \begin{subfigure}[t]{0.48\textwidth}
    \includegraphics[width=\linewidth]{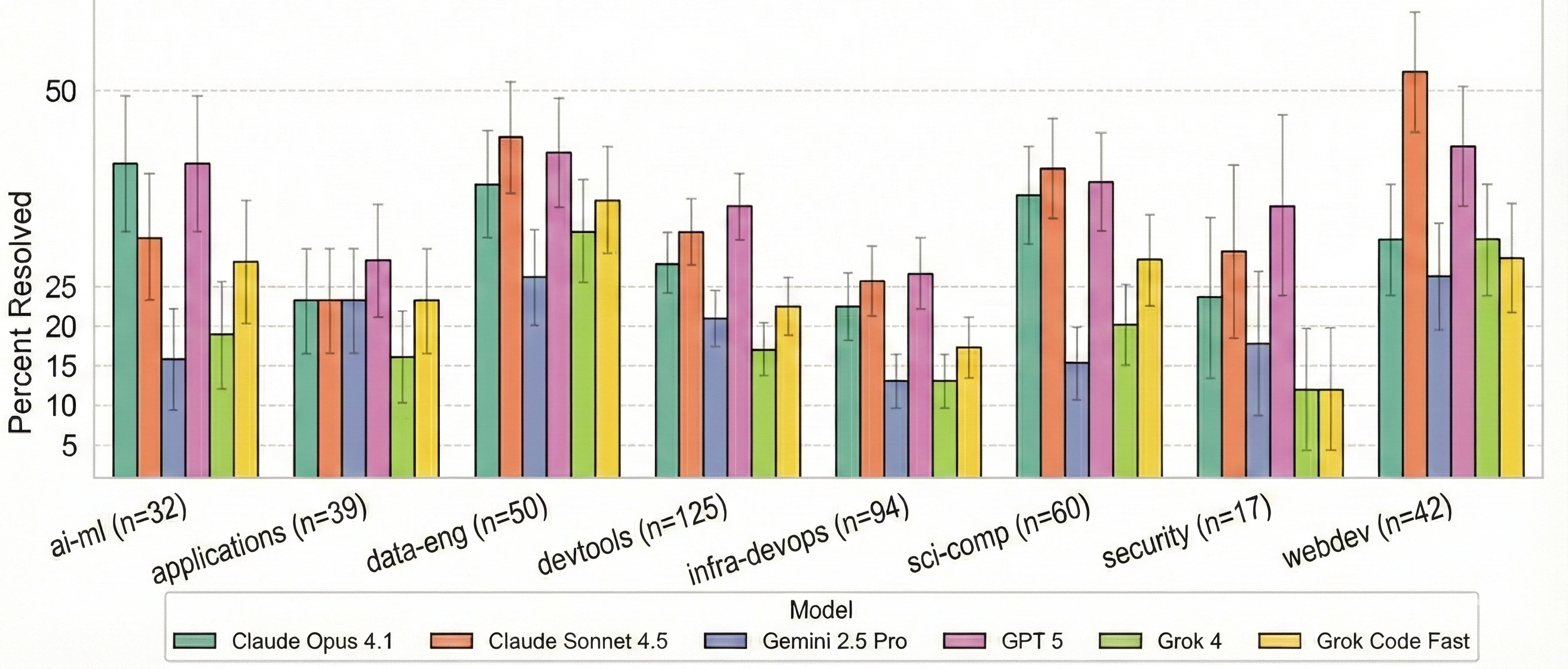} 
    \caption{Percentage of resolved repositories by type.}
    \label{fig:percentage-resolved-repos}
  \end{subfigure}\hfill
  \begin{subfigure}[t]{0.48\textwidth}
    \includegraphics[width=\linewidth]{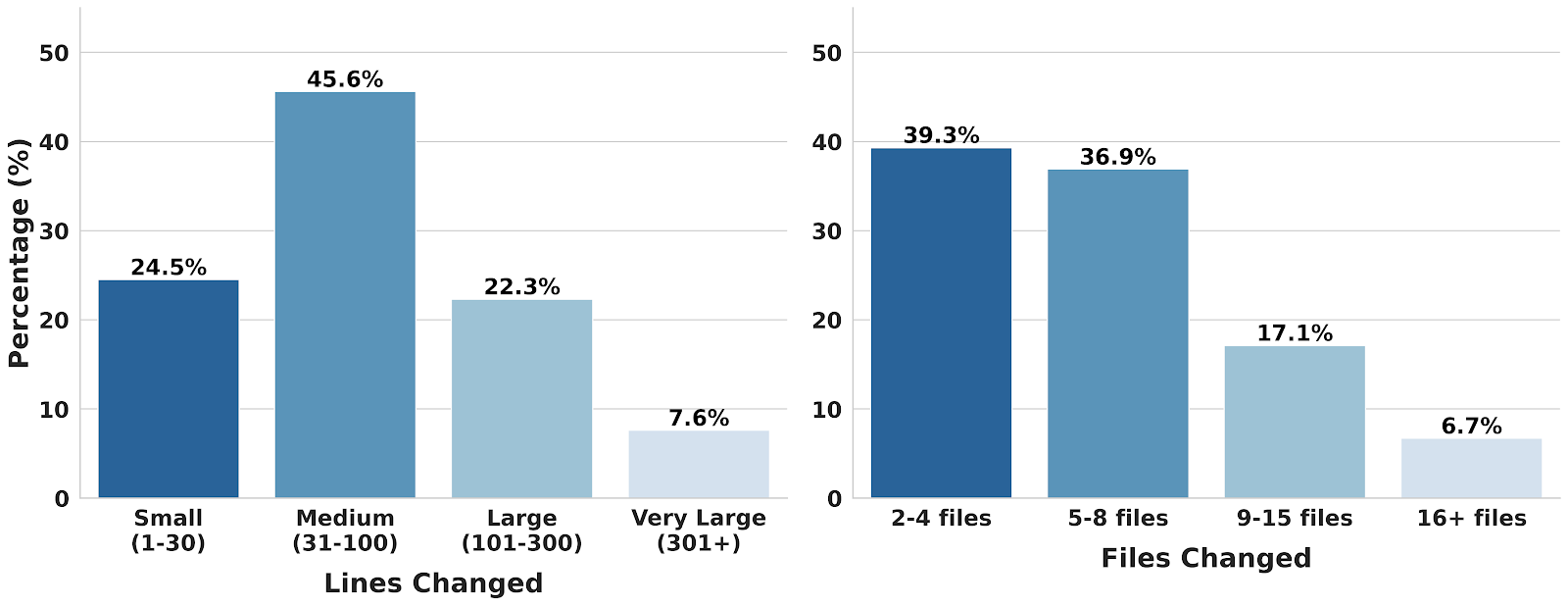}
    \caption{Difficulty distribution.}
    \label{fig:difficulty-distribution}
  \end{subfigure}
  \caption{An overall view of the dataset.}
  \label{fig:pair}
\end{figure}

\paragraph{Dataset Distribution} The final dataset covers \textbf{3,971 unique repositories}, an increase of two orders of magnitude over the 12 repositories in the original SWE-bench. This shift minimizes the risk of overfitting to specific project coding styles. A taxonomy analysis on a random subset of 488 instances confirmed the dataset's representative scope across the open-source distribution, covering diverse domains such as \textbf{DevTools (27.1\%)}, \textbf{Infra/DevOps (18.5\%)}, \textbf{Scientific Computing (12.9\%)}, and \textbf{Data Engineering (10.7\%)}, with the long tail covering AI/ML, Blockchain, and Embedded Systems. Furthermore, the evaluation of resolved rates, broken down by repository domain (as detailed in Figure \ref{fig:percentage-resolved-repos}), showed significant differential performance among the models. This suggests that the inherent diversity of the benchmark serves as a high-fidelity diagnostic tool, moving beyond single-score metrics to reveal the specific strengths and weaknesses of advanced code-generation models. 

\paragraph{Bug vs Feature Coverage} To quantify task types, we manually annotated a random sample of 488 instances. At the code level, 61.1\% of changes are primarily bug fixes, 30.7\% implement new features, with the remainder covering refactors, performance improvements, and other maintenance \ref{app:code-type}. At the issue level, 56.1\% of linked issues are bug reports and 38.5\% are feature requests \ref{app:issue-type}. This substantially increases feature‑request coverage compared to SWE‑bench, which contains only about 9\% feature‑request issues.

\paragraph{Difficulty Distribution} The SWE-Bench++ dataset has a balanced distribution of task sizes by lines changed: 24.5\% small (1--30), 45.6\% medium (31--100), 22.3\% large (101--300), and 7.6\% very large (301+). Files changed also show breadth: 39.3\% involve 2--4 files, 36.9\% 5--8 files, 17.1\% 9--15 files, and 6.7\% 16+ files. The dataset includes challenging instances (12.2\% with 200+ lines and 17.2\% with 10+ files), ensuring a comprehensive evaluation framework covering quick fixes to large-scale refactors.

\subsection{EVALUATING LLM AGENTS ON SWE-Bench++}
\label{sec:baseline-eval}

To establish performance baselines, we benchmarked leading LLM agents on a 1,782‑instance subset of SWE‑Bench++ (a verified cross‑lingual sample). We construct this subset via stratified random sampling by language, drawing a fixed number of instances from each language slice to preserve diversity while keeping evaluation cost manageable. This yields roughly 100–280 instances per language (full breakdown in Appendix \ref{app:sampling}). As shown in Table \ref{tab:leaderboard}, the benchmark presents a significant challenge even to frontier models.

\begin{itemize}
  \item \textbf{Performance Hierarchy:} The gap between the top-performing model (\textbf{36.20\%}) and mid-tier models (low-20s) highlights the difficulty of the benchmark.
  \item \textbf{Language Disparity:} Models generally perform stronger on Python and Java, likely due to the preponderance of these languages in pre-training corpora. Lower performance on Go and C++ highlights the need for multilingual benchmarks to drive progress in less represented languages.
\end{itemize}

\begin{table}[htbp]
\centering
\caption{Leaderboard of model performance on the SWE-Bench++ 1,782-instance subset using the SWE-Agent framework \citep{jimenez2024swebench}.}
\label{tab:leaderboard}

\resizebox{\linewidth}{!}{
    \setlength{\tabcolsep}{3pt} 
    \begin{tabular}{>{\raggedright\arraybackslash}p{4cm}cccccccccc}
    \toprule
     & \textbf{Overall} & \multicolumn{9}{c}{\textbf{Per-language pass@10}} \\
     \cmidrule(lr){3-11}
    \textbf{Model} & \textit{(SWE-Bench++ 1,782‑instance subset)} & \textbf{Py} & \textbf{Java} & \textbf{JS/TS} & \textbf{Rust} & \textbf{C/C++} & \textbf{Go} & \textbf{PHP} & \textbf{Ruby} & \textbf{C\#} \\
    \midrule
    gpt-5-2025-08-07 & 34.57\% & \textbf{43.57\%} & \textbf{41.84\%} & 33.67\% & 22.86\% & 30.81\% & 24.00\% & 42.90\% & 40.50\% & 39.00\% \\
    claude-sonnet-4.5 & \textbf{36.20\%} & 34.29\% & 39.80\% & \textbf{34.69\%} & 22.86\% & \textbf{57.30\%} & \textbf{28.00\%} & 42.90\% & \textbf{53.00\%} & \textbf{55.00\%} \\
    claude-opus-4.1 & 32.38\% & 37.14\% & 33.67\% & 27.55\% & 14.29\% & 48.11\% & 19.00\% & 42.90\% & 40.50\% & 43.50\% \\
    claude-sonnet-4-20250514 & 31.09\% & 36.43\% & 31.63\% & 26.53\% & 17.14\% & 36.22\% & 20.00\% & 35.70\% & 37.50\% & 42.50\% \\
    xai/grok-code-fast-1 & 30.42\% & 36.43\% & \textbf{41.84\%} & 31.63\% & 11.43\% & 10.27\% & \textbf{28.00\%} & \textbf{50.00\%} & 35.00\% & 42.00\% \\
    xai/grok-4-0709 & 25.52\% & 34.29\% & 38.78\% & 29.59\% & 22.86\% & 7.03\% & 17.00\% & 42.90\% & 34.50\% & 39.00\% \\
    gemini/gemini-2.5-pro & 24.92\% & 20.00\% & 28.57\% & 19.39\% & 8.57\% & 28.11\% & 13.00\% & \textbf{50.00\%} & 35.50\% & 34.00\% \\
    qwen3-coder & 24.19\% & 16.43\% & 31.63\% & 21.43\% & 14.29\% & 9.19\% & 19.00\% & 35.70\% & 39.50\% & 36.50\% \\
    gpt-4o & 16.89\% & 10.71\% & 12.24\% & 5.10\% & 5.71\% & 4.32\% & 9.00\% & 28.60\% & 13.00\% & 9.00\% \\
    \bottomrule
    \end{tabular}
}
\end{table}

\subsection{Fine-Tuning Experiments}
\label{sec:fine-tuning-experiments}

To validate the efficacy of SWE-Bench++ as a source of high-quality training data, we conducted a controlled fine-tuning study using the \textbf{Qwen2.5-Coder 7B and 32B} variants \citep{hui2024qwen25codertechnicalreport}. Our objective was to assess whether adding a small volume of real-world, agentic trajectories from \emph{SWE-Bench++} could improve performance on out-of-distribution polyglot tasks compared to purely synthetic baselines.

\subsubsection{Experimental Setup}
We constructed distinct data mixtures to isolate the impact of trajectory source, diversity and scale. 
\begin{enumerate}
  \item \textbf{Experiment 1 (Baseline):} We establish a baseline using the 5{,}016 synthetic trajectories from SWE-Smith. The dataset used here serves as the foundation for the SWE-Agent-LM models (7B and 32B denoted as Experiment 1a and 1b respectively).
  \item \textbf{Experiment 2 (SWE-Bench++ Density):} We augment the Baseline with \textbf{179 curated trajectories} sourced from \textbf{44 GitHub issues} via \emph{SWE-Bench++}. This mixture tests whether having multiple solution paths for the same issues helps more than adding new issues.
  \item \textbf{Experiment 3 (SWE-Bench++ Diversity):} We augment the Baseline with \textbf{145 curated trajectories} sourced from \textbf{145 unique GitHub issues}. This mixture tests the value of maximizing issue and repository variety (higher diversity) with a comparable data volume.
  \item \textbf{Experiment 4 (SWE-Bench++ Data-Scaling):} We evaluate data scaling laws by iteratively augmenting the Baseline with \textbf{200}, \textbf{400}, and \textbf{800} curated trajectories (denoted as \emph{SWE-Bench++ Data-Scaling-1}, \emph{SWE-Bench++ Data-Scaling-2}, and \emph{SWE-Bench++ Data-Scaling-3}). Distinct from Experiments 2 and 3, the trajectories here utilize a hybrid human review strategy where 40\% of the added data is purely synthetic (filtered only for passing the harness but with no human QA) while the remainder undergoes human review.
  \item \textbf{Experiment 5 (SWE-Bench++ Model-Scaling):} We replicate the data scaling variations from Experiment 4 using the \textbf{Qwen2.5-Coder-32B} model to verify that the performance gains hold also for models of much larger sizes. Note that for the \emph{SWE-Bench++ Data-Scaling-3 variation} (+800 SWE-Bench++ data) we scaled the SWE-Smith baseline to \textbf{10{,}032 trajectories}. This ensures a robust ratio between the specialized SWE-Bench++ data and the general baseline, preventing the larger model from overfitting to the specific fine-tuning tasks at the expense of generalization (catastrophic forgetting).
\end{enumerate}

\textbf{Evaluation Protocol:} We evaluate the resulting checkpoints on \textbf{SWE-Bench Multilingual} \citep{yang2025swesmithscalingdatasoftware}, a rigorous benchmark comprising 300 tasks across 42 repositories and 9 programming languages. This benchmark contains \textbf{no Python tasks}, serving as a strict test of cross-lingual generalization. Furthermore, we ensured zero overlap between the repositories used in our \emph{SWE-Bench++} training sets and the \emph{SWE-bench Multilingual} test set.

\textbf{Ablation Logic:} \textbf{Experiment 1} functions as a reproduction of the SWE-Agent-LM performance profile, utilizing a standard fine-tuning setup in the absence of open-source configurations. \textbf{Experiments 2 and 3} disentangle the value of repository variety versus data volume, using raw, unverified trajectories to establish a strict lower-bound for data utility. \textbf{Experiments 4 and 5} confirm that gains are robust across data and model sizes; crucially, the inclusion of 40\% unreviewed data verifies that our high signal-to-noise sourcing allows for scaling without being bottlenecked by manual review costs.

\subsubsection{Fine-Tuning Experiment Details}
\label{fine-tuning-experiment-details-main}

Our fine-tuning setup closely follows SWE-Smith. We use a learning rate of 5e‑5, up to 3 epochs, and a maximum context length of 32,768, and we run training with MS‑Swift on 8× NVIDIA H200 144G GPUs. We also follow the same XML data conversion strategy and the rejection sampling fine-tuning process. A detailed list of all the used hyper-parameters can be found in Appendix \ref{fine-tuning-experiment-details}.

\subsubsection{Results}
The results of the experiment can be seen in the table below:

\begin{table}[htbp]
\centering
\small
\begin{tabular}{lclcc}
\hline
\textbf{Model Size} & \textbf{Experiment} & \textbf{Fine-Tuning Mixture} & \textbf{Performance (pass@1)} & \textbf{Diff CI* (95\%)} \\
\hline
Qwen2.5-Coder-7B & 0 & Off-the-shelf & 0 / 300 & -- \\
 & 1 & SWE-Smith 5k & 5 / 300 & (+1.0, +10.0) \\
 & 2 & SWE-Bench++ Density & 7 / 300 & (+0.0, +5.0) \\
 & 3 & SWE-Bench++ Diversity & \textbf{11 / 300} & (+1.0, +8.0) \\
 & 4 & SWE-Bench++ Data-Scaling-1 & 6 / 300 & -- \\
 &   & SWE-Bench++ Data-Scaling-2 & 16 / 300 & (+4.0, +16.0) \\
 &   & SWE-Bench++ Data-Scaling-3 & \textbf{20 / 300} & (+1.0, +8.0) \\
\hline
Qwen2.5-Coder-32B & 0 & Off-the-shelf & 4 / 300 & -- \\
 & 1 & SWE-Smith 5k & 12 / 300 & (+3.0, +14.0) \\
 & 5 & SWE-Bench++ Data-Scaling-1 & 17 / 300 & (+1.0, +10.0) \\
 &   & SWE-Bench++ Data-Scaling-2 & 21 / 300 & (+1.0, +8.0) \\
 &   & SWE-Bench++ Data-Scaling-3* ** & \textbf{25 / 300} & (+1.0, +8.0) \\
\hline
\end{tabular}
\caption{Fine-tuning results on SWE-bench Multilingual.}
\label{tab:fine_tuning_results}
\end{table}

{\footnotesize
\emph{*} 95\% confidence interval of the difference between the pass@1 performance of the current and the previous row.\\
\emph{**} this variation uses 10{,}032 SWE-Smith data in the baseline mix instead of 5{,}016 in all the other experiments.\\
}

Incorporating just 145 SWE-Bench++ trajectories (i.e., 2.8\% of the mix) increased the baseline performance (from 5/300 to 11/300) and more than doubled the number of valid patches, demonstrating the critical value of high-diversity, ``hard'' multilingual samples. While the improvement from solution density (Exp 2) is small and not statistically distinguishable from the SWE‑Smith baseline (95\% CI: +0.0 to +5.0), the improvement from repository variety (Exp 3) presents a robust, statistically significant signal (95\% CI: +1.0 to +8.0). \textbf{These gains scale monotonically with data volume}: increasing the SWE-Bench++ subset to 800 trajectories quadrupled the 7B baseline score, a trend that transferred robustly to the 32B model which reached a peak performance of 25/300.

\section{Limitations and future work}

While SWE-Bench++ scales via automation, execution-based verification remains a proxy for correctness and cannot capture aspects such as code maintainability or algorithmic efficiency. Furthermore, our \emph{State-Differential Oracle} is bound by the quality of the original developer-written test suites; sparse tests may strictly allow agents to produce patches that technically pass but fail human review. While our current pipeline utilizes an LLM-Judge for semantic alignment, approximating professional maintainer standards remains a challenge. Future iterations could explore 1) scalable Human-in-the-Loop curation, leveraging community feedback or crowdsourced validation to bridge the gap between functional correctness and true patch acceptability; 2) multi-modal verification to support UI-centric or frontend-heavy tasks.

Our “living benchmark” design enables temporally separated, contamination‑aware evaluation by filtering instances based on PR creation time relative to a model’s pre‑training cutoff. In practice, we approximate contamination control by evaluating each model only on instances whose PR timestamps fall strictly after its published cutoff date. While temporal separation is a strong heuristic rather than a formal guarantee—as code may still appear in other pre‑training sources—the framework’s primary defense is its capacity for continuous regeneration. Unlike static datasets, SWE‑Bench++ continuously ingests fresh pull requests, enabling dynamic evaluation sets that extend beyond any fixed training horizon.

\section{Conclusion}
Real-world software development is heterogeneous and evolving—properties that static, manually curated benchmarks struggle to capture. While previous efforts laid the groundwork for repository-level evaluation, manual bottlenecks restricted them to a negligible fraction of the ecosystem. SWE-Bench++ addresses these limitations not merely through automation, but through a constrained neural synthesis framework. By integrating template-guided environment scaffolding, state-differential oracle extraction, and adaptive log parsing, our approach systematically recovers complex tasks—including feature requests—across heterogeneous build systems that prior methods discard. Furthermore, our hint-guided trajectory synthesis transforms these tasks into a vital training resource that improves model performance on an external multilingual benchmark. By maintaining a living benchmark of fresh instances, SWE-Bench++ minimizes data contamination risks while driving the development of models that generalize across languages and build systems.

\section{Acknowledgments}
We thank David Wei, Mahesh Joshi, Yuzhao Ni, Ivan Kuznetsov, Manas Sambare, Mithil Poojary, Ashni Sheth for their insightful discussions and
valuable feedback. We extend our deepest thanks to all
annotators for their tremendous effort and contributions to this project.

\bibliography{main} 
\bibliographystyle{main}

\appendix


\clearpage            
\appendix

\section{Sample Input and Output in §3.2.2}

\subsection{Phase 1: Sample Input and Output}

\textbf{Input.}

\label{app:build_engineer_prompt}

\begin{promptbox}{System Prompt: Expert Build Engineer}

\noindent \textbf{Role:} Expert Build Engineer AI \\
\textbf{Task:} Analyze a GitHub pull request and generate a JSON configuration object to populate the Dockerfile template. \\
\textbf{Goal:} Populate \texttt{pre\_install}, \texttt{build}, \texttt{test\_cmd}, and \texttt{log\_parser\_name} fields.

\vspace{0.5em}
\hrule
\vspace{0.5em}

\textbf{Core Reasoning Steps:}

\begin{enumerate}
    \item \textbf{Repository Inspection:}
    \begin{itemize}
        \item \textbf{Submodules:} Detect \texttt{.gitmodules} and ensure recursive initialization.
        \item \textbf{Build System Detection:} Identify primary build tools by file existence (e.g., \texttt{CMakeLists.txt} $\to$ CMake, \texttt{pom.xml} $\to$ Maven).
        \item \textbf{Compiler Standard:} Parse configuration files to detect language standards (e.g., \texttt{CMAKE\_CXX\_STANDARD}, \texttt{sourceCompatibility}).
    \end{itemize}

    \item \textbf{Dependency Resolution:}
    \begin{itemize}
        \item \textbf{System Dependencies:} Scan manifests/imports for system packages (e.g., \texttt{apt-get install}).
        \item \textbf{Package Managers:} Detect/invoke language-specific managers (e.g., \texttt{vcpkg}, \texttt{pip}, \texttt{npm}, \texttt{cargo}).
    \end{itemize}

    \item \textbf{Build Strategy:}
    \begin{itemize}
        \item Generate commands to clean, configure, and build the project.
        \item \textbf{Constraint:} Ensure build artifacts persist in dedicated directories.
        \item \textbf{Constraint:} Enforce parallel compilation (e.g., \texttt{-j\$(nproc)}) for efficiency.
    \end{itemize}

    \item \textbf{Test Execution Logic (Crucial):}
    \begin{itemize}
        \item \textbf{Scope filtering:} Map \textit{modified files} in the PR to specific test targets (e.g., \texttt{pytest tests/test\_mod.py}) if possible.
        \item \textbf{Output Formatting:} Ensure the test runner emits machine-readable logs (XML/JSON) or verbose stdout.
    \end{itemize}

    \item \textbf{Log Parser Selection:}
    \begin{itemize}
        \item Analyze test framework imports (e.g., \texttt{import pytest}, \texttt{\#include <gtest.h>}) to select parser:
        \item \texttt{"googletest" | "pytest" | "maven" | ... | "agentic"}
    \end{itemize}
\end{enumerate}

\vspace{0.5em}
\textbf{Output Specification (JSON):}

\begin{lstlisting}[language=json, frame=none, backgroundcolor=\color{white}]
{
  "pre_install": ["list", "of", "system", "commands"],
  "build": ["list", "of", "build", "commands"],
  "test_cmd": "single comprehensive test command",
  "log_parser_name": "selected_parser_id",
}
\end{lstlisting}

\end{promptbox}

\newpage

\textbf{Output.}
\begin{verbatim}
{
  "docker_specs": {
    "java_version": "17"
  },
  "build": [
    "mvn clean install -B -DskipTests"
  ],
  "test_cmd": "mvn test -B -pl project-core,project-api -am",
  "log_parser_name": "maven"
}
\end{verbatim}

\subsection{Phase 2: Sample Input and Output}

\textbf{Input.}

\begin{promptbox}{System Prompt: Error Recovery \& Fix}

\textbf{Context:} A previous attempt to generate a build configuration failed during execution.

\noindent \textbf{Task:} Analyze the provided error log and the original JSON configuration to identify the root cause. Based on your analysis, provide a revised JSON object that corrects the error.

\vspace{0.5em}
\hrule
\vspace{0.5em}

\textbf{Original Failing Configuration (JSON)}
\begin{lstlisting}[language=json, frame=none, backgroundcolor=\color{white}]
{
  "docker_specs": { "java_version": "17" },
  "build": [ "mvn clean install -B -DskipTests" ],
  "test_cmd": "mvn test -B -pl project-aggregator -am",
  "log_parser_name": "maven"
}
\end{lstlisting}

\vspace{0.5em}
\textbf{Execution Error Log}
\begin{lstlisting}[language=logoutput]
+ mvn test -B -pl project-aggregator -am
[INFO] Scanning for projects...
[ERROR] Could not find the selected project 'project-aggregator' in the reactor.
[ERROR] Available projects: project-parent, project-core, project-api
...
\end{lstlisting}

\end{promptbox}

\textbf{Output.}
\begin{verbatim}
{
  "docker_specs": {
    "java_version": "17"
  },
  "build": [
    "mvn clean install -B -DskipTests"
  ],
  "test_cmd": "mvn test -B -pl project-core,project-api -am",
  "log_parser_name": "maven"
}
\end{verbatim}

\newpage

\subsection{Phase 3: Sample Input and Output}

\textbf{Input.}

\begin{promptbox}[unbreakable]{System Prompt: Test Execution Verification}

\noindent \textbf{Role:} Test Execution Validator \\
\textbf{Task:} Analyze the terminal output from a test command to verify if the execution was valid, complete, and semantically parsable. \\
\textbf{Goal:} Distinguish between a \textit{valid test run} (where tests may fail due to code bugs) and a \textit{broken environment} (where the runner crashes or fails to start).

\vspace{0.5em}
\hrule
\vspace{0.5em}

\textbf{Core Reasoning Steps:}

\begin{enumerate}
    \item \textbf{Execution Integrity:} Check if the test runner successfully started and completed its process without crashing or exiting early due to segmentation faults or system errors.
    
    \item \textbf{Output Completeness:} Verify that specific test names or IDs were printed to stdout (e.g., \texttt{test\_login PASSED}). This is a prerequisite for the subsequent Log Parsing stage.
    
    \item \textbf{Scope Verification:} Confirm that the test suite ran the intended scope (all tests or the targeted subset), rather than silently exiting after a single submodule error.
    
    \item \textbf{Failure Classification (Crucial):}
    \begin{itemize}
        \item \textbf{Acceptable (Success):} Tests marked as \texttt{FAIL}, \texttt{SKIP}, or \texttt{XFAIL} due to logic bugs, missing credentials, or platform constraints (e.g., GPU required). These indicate a working environment. $\to$ \texttt{success: true}
        \item \textbf{Unacceptable (Failure):} Failures due to missing system libraries (\texttt{ImportError}, \texttt{ModuleNotFound}), toolchain crashes, or syntax errors in the test harness itself. These indicate a broken environment that requires rebuilding. $\to$ \texttt{success: false}
    \end{itemize}
\end{enumerate}

\vspace{0.5em}
\textbf{Output Specification (JSON):}

\begin{lstlisting}[style=jsonWithComments]
{
  "success": true, // Set to true if environment is healthy
  "reason": "Explanation if success is false",
  "error_message": "Extracted environment error trace",
  "details": {
    "testCommandExecuted": true,
    "testNamesPrinted": true,
    "allTestsRan": true
  }
}
\end{lstlisting}

\end{promptbox}

\textbf{Output.}
\begin{verbatim}
{
  "success": true,
  "reason": "",
  "error_message": "",
  "details": {
    "testCommandExecuted": true,
    "testNamesPrinted": true,
    "mostTestsRan": true
  }
}
\end{verbatim}

\clearpage

\section{Standard Log Parser Utilized}
\label{sec:log-parsers}

\begin{longtable}{>{\raggedright\arraybackslash}p{0.22\textwidth} >{\raggedright\arraybackslash}p{0.73\textwidth}}
\toprule
\textbf{Language} & \textbf{Log parser} \\
\midrule
Python &
\texttt{pytest} — Pytest; \texttt{django} — Django test runner \\
\addlinespace
JavaScript / TypeScript &
\texttt{vitest} — Vitest; \texttt{jest} — Jest; \texttt{mocha} — Mocha; \texttt{karma} — Karma; \texttt{tap} — Test Anything Protocol \\
\addlinespace
Java &
\texttt{maven} — Maven Surefire; \texttt{gradle} — Gradle; \texttt{ant} — Ant \\
\addlinespace
Go &
\texttt{gotest} — standard \texttt{go test} \\
\addlinespace
Rust &
\texttt{cargo} — standard \texttt{cargo test} \\
\addlinespace
Ruby &
\texttt{rubyunit}; \texttt{minitest}; \texttt{rspec} — with JSON output transformation; \texttt{cucumber}; \texttt{tap} \\
\addlinespace
PHP &
\texttt{phpunit} \\
\addlinespace
C/C++ &
\texttt{doctest} — XML doctest; \texttt{googletest}; \texttt{catch2}; \texttt{tap} \\
\addlinespace
C\# &
\texttt{NUnit}; \texttt{XUnit}; \texttt{MSTest} \\
\bottomrule
\end{longtable}

\section{LLM-Judge for Layer 3 Semantic Alignment in §3.4}
\label{app:llm-judge}

\subsection{Quality Analysis Metrics}

To systematically assess PR quality, SWE-Bench++ computes two complementary scores—\emph{issue\_clarity} and \emph{test\_to\_issue\_alignment}—each ranging from 0 (best) to 3 (worst).

\begin{itemize}[leftmargin=2em]
  \item \textbf{Presence of Success Criteria:} Explicit expected behavior or acceptance criteria; missing or vague criteria increase the score.
  \item \textbf{Specificity of Problem Description:} Concrete, unambiguous instructions reduce the score; motivational-only content, screenshots, or external links increase it.
  \item \textbf{Contextual Completeness:} Steps to reproduce, code snippets, or stack traces reduce the score; absence of actionable information leads to the highest score.
\end{itemize}

\subsection{Quantifying Test-to-Issue Alignment}

\begin{itemize}[leftmargin=2em]
  \item \textbf{Core Behavior Coverage:} Tests should exercise the main functionality or bug; poor coverage yields higher scores (2--3).
  \item \textbf{False Negatives (Correct Solutions Rejected):} Tests too narrow; typically addressed by adding cases (e.g., score 1).
  \item \textbf{False Positives (Incorrect Solutions Accepted):} Missing core coverage; requires extending/modifying tests (penalized 2--3).
\end{itemize}

\subsection{Performance by Human Verification}

Automated reviewers—intended as junior substitutes—already achieve precision close to senior/lead annotators (0.93–0.95 vs. 0.964/0.963). Recall varies substantially: Claude‑Sonnet‑4 (0.921) and Gemini‑2.5‑Pro (0.884) are closest to human recall (0.991/0.963), whereas GPT‑5 is markedly more conservative (0.370 recall, FNR 0.63). This conservatism yields the lowest false positive rate (0.154) but at the cost of missing many problematic instances, while the higher‑recall automated reviewers trade off with elevated false positive rates relative to humans (0.423–0.615 vs. 0.308).

\begin{table}[t]
\centering
\begin{tabular}{lcccc}
\toprule
Reviewer & Precision & Recall & False Pos. Rate & False Neg. Rate \\
\midrule
Auto Reviewer (Claude-Sonnet-4)   & 0.926 & 0.921 & 0.615 & 0.079 \\
Auto Reviewer (Gemini-2.5-Pro)    & 0.946 & 0.884 & 0.423 & 0.116 \\
Auto Reviewer (GPT-5)             & 0.952 & 0.370 & 0.154 & 0.630 \\
Senior (qa\_trainer3)             & 0.964 & 0.991 & 0.308 & 0.009 \\
Lead (calibrator)                 & 0.963 & 0.963 & 0.308 & 0.037 \\
\bottomrule
\end{tabular}
\caption{Performance comparison of automated LLM-based reviewers and human annotators.}
\label{tab:auto-reviewers}
\end{table}

\section{Hints Generation Process for Layer 3 Semantic Alignment in §3.4}
\label{app:hints}

\subsection{Predicting Whether Hints Are Needed (\texttt{is\_hint\_needed})}
The module predicts whether a hint is necessary (\texttt{is\_hint\_needed}=1 indicates needed). Decision signals:
\begin{itemize}[leftmargin=2em]
  \item \textbf{Build Logs:} Detect build failures in “Before” logs via log parsing.
  \item \textbf{Golden Rules:} Identify elements like new function signatures to prevent avoidable failures (e.g., \texttt{new\_function} vs.\ \texttt{newFunction}) via AST/regex.
  \item \textbf{LLM Judgment:} Few-shot LLM assesses whether contextual hints are required.
\end{itemize}
Evaluated on 243 senior-labeled instances; accuracy was 94.6\%.

\subsection{Generating Hint Values}
Generate minimal \texttt{hint\_value} content focusing on essentials (e.g., critical signatures), combining golden rules with LLM judgment to maximize fairness while avoiding biasing context.

\section{\texttt{code\_type} and \texttt{issue\_type} Distributions in §4.1}

\subsection{Code Type (Primary)}
\label{app:code-type}
\begin{table}[h]
\centering
\begin{tabular}{lrr}
\toprule
\textbf{code\_type\_primary} & \textbf{count} & \textbf{percentage} \\
\midrule
bug-fix            & 298 & 61.0656 \\
feature            & 150 & 30.7377 \\
refactor           & 22  & 4.5082  \\
performance        & 12  & 2.4590  \\
unknown            & 3   & 0.6148  \\
dependency-update  & 1   & 0.2049  \\
build-ci           & 1   & 0.2049  \\
\bottomrule
\end{tabular}
\end{table}

\subsection{Issue Type (Primary)}
\label{app:issue-type}
\begin{table}[h]
\centering
\begin{tabular}{lrr}
\toprule
\textbf{issue\_type\_primary} & \textbf{count} & \textbf{percentage} \\
\midrule
bug-report       & 274 & 56.1475 \\
feature-request  & 188 & 38.5246 \\
performance-issue& 12  & 2.4590  \\
chore            & 10  & 2.0492  \\
unknown          & 2   & 0.4098  \\
\bottomrule
\end{tabular}
\end{table}

\subsection{Language breakdown for the 1,782 samples}
\label{app:sampling}
\begin{table}[h]
\centering
\begin{tabular}{lrr}
\toprule
\textbf{language} & \textbf{sample count} \\
\midrule
Python         & 280    \\
JavaScript     & 100  \\
TypeScript     & 96    \\
Java           & 196   \\
Go             & 200   \\
Rust           & 175  \\
Ruby           & 200   \\
PHP            & 100   \\
C/C++          & 185  \\
C\#            & 250  \\

\bottomrule
\end{tabular}
\end{table}

\section{Trajectory Generation Process in §4.3}
\label{app:trajectory}

\subsection{Data Preparation}

We select \textbf{SOTA model-breaking} issues as the basis for curation, leveraging \textbf{tailored Docker environments} produced earlier in the pipeline so the agent does not need to install the environment. The scaffold supports multiple underlying LLMs with SWE-Agent. This choice is motivated by the goal of increasing \textbf{trajectory diversity}---as different models induce distinct solution paths. 

\subsection{Trajectory Generation and Selection}

We produce successful trajectories by engineering the system/user prompts and injecting \textbf{issue-tailored hints} that guide the agent’s exploration and problem solving. Successful trajectories are identified using our extended SWE-Bench evaluation harness: a trajectory is marked as successful if its submitted solution yields successful outcomes on both \textbf{P2P} (pass-to-pass) and \textbf{F2P} (fail-to-pass) tests. Hints are crucial---combined with multiple attempts, they raise the passing rate from approximately 0\% to $\sim$70\% in our setting---enabling effective demonstration generation on SOTA-model-breaking cases.

We curate agentic trajectories that successfully resolve real-world GitHub issues, encompassing a diverse array of task types including bug fixes, feature requests, and refactoring. This approach contrasts fundamentally with synthetic pipelines like \textbf{SWE-Smith} \citep{yang2025swesmithscalingdatasoftware}, which typically construct artificial problem constraints and focus predominantly on regression bugs. 

By leveraging the \textit{SWE-Bench++} automated pipeline and the novel hint injection strategy (Section 3.5), we are able to harvest these high-fidelity traces to SOTA ``model breaking'' GitHub issues at scale without reliance on pre-existing model solutions. 

We utilize \textbf{SWE-Agent} \citep{jimenez2024swebench} as the agentic scaffold. Each trajectory is recorded as a structured, interleaved sequence of:

\begin{enumerate}
    \item \textbf{Thought:} Model-generated reasoning traces (Chain-of-Thought).
    \item \textbf{Action:} Executable tool invocations (e.g., \texttt{edit\_file}, \texttt{run\_test}).
    \item \textbf{Observation:} Verbatim feedback from the environment (stdout/stderr).
\end{enumerate}

This \texttt{(Thought, Action, Observation)} representation captures the dynamics of iterative problem-solving, demonstrating how an agent explores a codebase, revises incorrect assumptions, and converges on a solution. This provides significantly richer supervision than the single-shot solution prompting used in the original SWE-bench setup \citep{jimenez2024swebench} or the static flows of ``agentless'' approaches~\citep{xia2025agentless}. These trajectories serve as high-quality demonstrations for standard fine-tuning regimes (e.g., SFT, DPO). 

\subsection{Automated Trajectory Quality Assurance}

The harness alone is not a sufficient measure of quality. For instance, truncated trajectories may still pass, and hints can induce over-reliance. We therefore apply a multi-stage, automated QA pipeline:

\begin{enumerate}
    \item \textbf{Structural validity:} remove trajectories that do not end with a final \textbf{submit} action.
    \item \textbf{Hint contamination control:} remove trajectories whose \textbf{actions} or \textbf{observations} contain hint-related keywords.
    \item \textbf{Thought regeneration:} prompt a language model to rewrite \textbf{thoughts} containing hint-related keywords (the prompt excludes the hints).
    \item \textbf{Automated judging:} use a language-model-based evaluator to retain only high-quality trajectories.
\end{enumerate}

The final automated stage (step 4 above) scores trajectories along four dimensions: (i) successful reproduction of the problem (for bug-fix cases), (ii) plausibility of the proposed solution, (iii) evidence of validation, and (iv) adherence to sound engineering practices.

\subsection{Human Review}

Trajectories that pass automated thresholds undergo \textbf{expert evaluation} (HITL -- human-in-the-loop). Reviewers detect non-trivial over-reliance on hints and systematic failure modes (e.g., a model consistently neglecting specific tools). They may regenerate thoughts, prune steps, or discard trajectories when warranted, focusing on quality aspects that cannot be reliably handled by algorithmic checks.

\vspace{0.5em}
\noindent \textbf{Outcome.} The result is a scalable, HITL-augmented curation process that yields agentic demonstrations solving \textbf{SOTA-model-breaking} issues. By encoding both reasoning and interaction dynamics, these trajectories are expressly designed for \textbf{fine-tuning} code-capable language models.

\section{Fine-Tuning Experiment Details in §4.3}
\label{fine-tuning-experiment-details}

As in the main paper, we follow the SWE-Smith setup \citep{yang2025swesmithscalingdatasoftware}. Key hyperparameters:
\begin{verbatim}
train_type full
torch_dtype bfloat16
num_train_epochs 3
per_device_train_batch_size 1
learning_rate 5e-05
weight_decay 0.01
lr_scheduler_type cosine
warmup_ratio 0.05
gradient_accumulation_steps 1 # 8 GPUs => effective batch-size = 8
max_length 32768
truncation_strategy 'right'
\end{verbatim}

\clearpage

\section{Case Studies in Failure Analysis of SOTA Models}
\label{sec:case-studies-failure-analysis}

\subsection{Case 1 — \texttt{yamllint}: Misinterpreting Raw vs.\ Preprocessed Tokens}

\textbf{Task Context}
\begin{itemize}[leftmargin=2em]
  \item Repository: \verb|adrienverge__yamllint-746| (\url{https://github.com/adrienverge/yamllint/pull/746})
  \item Language: Python
  \item Domain: Developer Tools
  \item Issue Type: Bug Fix (bug-report)
\end{itemize}

\textbf{Issue description from PR}
\begin{quote}
On double-quoted multiline strings, quotes aren't needed if lines are broken on spaces, e.g.: \texttt{multiline: "this is a sentence   cut into words"}.
But quotes are needed when at least one line ends with a backslash (\texttt{\textbackslash}), meaning the next spaces should be removed:
\texttt{multiline: "https://example.com/a/very/very\textbackslash\quad/very/very/long/URL"}.
\end{quote}

\textbf{SOTA Model Erroneous Reasoning Trace (claude-sonnet-4.5)}
\begin{quote}\small
``The issue is in the \texttt{\_quotes\_are\_needed} function. When this function checks if quotes are needed, it creates a test YAML string like \texttt{key: <string>} and tries to parse it. The problem is that when the original string was a multiline quoted string, the \texttt{token.value} contains the final parsed value (without backslashes), not the original form from the file. Let me create a test to better understand how tokens work with multiline strings: \dots''
\end{quote}

The model assumed tokens were preprocessed, leading to:
\begin{verbatim}
# Quotes needed for multiline strings with line continuations (backslashes)
if is_multiline_string:
    return True
\end{verbatim}

\textbf{Diagnosis by Our Framework} \\
\texttt{yamllint} reads raw token buffers; backslashes are preserved. A helper inspects raw buffers:
\begin{verbatim}
def _has_backslash_on_at_least_one_line_ending(token):
    if token.start_mark.line == token.end_mark.line:
        return False
    buffer = token.start_mark.buffer[
        token.start_mark.index + 1:token.end_mark.index - 1]
    return '\\\n' in buffer or '\\\r\n' in buffer
\end{verbatim}

\subsection{Case 2 — \texttt{changedetection.io}: Mishandling Multiline Regex and Zero-Width Matches}

\textbf{Task Context}
\begin{itemize}[leftmargin=2em]
  \item Repository: \verb|dgtlmoon__changedetection.io-2889| (\url{https://github.com/dgtlmoon/changedetection.io/pull/2889})
  \item Language: Python
  \item Domain: Applications
  \item Issue Type: Feature Request
\end{itemize}

\textbf{Issue description from PR}
\begin{quote}
Support multiline regex in text filtering. The current function iterates per line:
\begin{verbatim}
for line in content.splitlines(keepends=True):
\end{verbatim}
Rework to use \texttt{re.finditer}/\texttt{re.findall} over the whole content.
\end{quote}

\textbf{SOTA Model Erroneous Reasoning Trace (Claude Sonnet 4.5)} \\
Tried span-overlap heuristics:
\begin{verbatim}
# Check if this line overlaps with any ignored range
for range_start, range_end in merged_ranges:
    if not (line_end <= range_start or line_start >= range_end):
        ignored_line_numbers.append(line_num)
        break
\end{verbatim}

Failed a unit test for zero-width multiline regex \verb|/^$/ms$|.

\textbf{Diagnosis by Our Framework}
\begin{enumerate}[leftmargin=2em]
  \item Evaluate regex over entire content with \texttt{re.MULTILINE} (and \texttt{re.DOTALL} as needed) using \texttt{re.finditer}.
  \item Map \texttt{match.start()} to line numbers by counting \texttt{\textbackslash n}.
  \item Avoid overlap heuristics; compute line indices from newline offsets.
\end{enumerate}
Zero-width spans break overlap checks; whole-content matching plus newline mapping fixes it.

\end{document}